\documentclass[aps,prc,floats,floatfix,preprint,superscriptaddress,nofootinbib]{revtex4}   
\def\be{\begin{eqnarray}}     
\def\ee{\end{eqnarray}}  
\usepackage{graphicx}  
  
\begin{document}  
  
\title{Combinatorial level densities by the real-time method}
 
\author{G.F.~Bertsch}
\affiliation{Institute for Nuclear Theory and Dept. of Physics, Box 351560, University 
of Washington, Seattle, Washington 98915, USA} 
\email{bertsch@uw.edu}
\author{L.M.~Robledo} 
\affiliation{Departamento de F\'\i sica Te\'orica, M\'odulo 15, Universidad Aut\'onoma de 
Madrid, E-28049 Madrid, Spain} 
 
\email{luis.robledo@uam.es} 
\homepage{http://gamma.ft.uam.es/robledo}

\begin{abstract} 
Levels densities of independent-particle Hamiltonians can be calculated
easily by using the real-time representation of the
evolution operator together with the fast Fourier transform.  We
describe the method and implement it with a set of Python programs.
Examples are provided for the total and partial levels densities 
of a heavy deformed nucleus ($^{164}$Dy).  The partial level densities
that may be calculated are the projected ones on neutron number, proton
number, azimuthal angular momentum, and parity.
\end{abstract}
\maketitle  
\section{introduction}  
  
Knowledge of nuclear level densities is important in 
nuclear reactions and decays, particularly for heavy nuclei.  The starting
point in the theory of level densities is the independent nucleon model, 
either from a shell model or a mean-field approximation.  A quantitative
theory requires a careful treatment of the interactions beyond mean field
\footnote{As examples of methods that can treat interactions in a realistic
way, we mention the interacting shell model approach\cite{se13} and 
the shell-model Monte Carlo approach\cite{al99}.},
but it is useful to have the independent particle level densities to
build from.  In principle that theory is very simple, needed only the 
single-particle energies to calculate the excitations.
Even so, the computational problem remains nontrivial.  The two well-known methods for treating
it are the statistical approach using the partition function,
and the combinatorial approach which individual particle-hole excitations
are counted.
The statistical approach via the partition function has two drawbacks.  One is 
that the calculated partition function must be transformed to an actual level density
by using the saddle-point approximation.  This is accurate when the level
density is high but is unsatisfactory as a complete solution. 
The other drawback is that partial level densities
(mostly associated with conserved quantum numbers) may be awkward
to extract.  The combinatorial approach does not require that the
level density be high, but it can also be awkward for writing codes
when it depends
on many quantum numbers that are to be exhibited explicitly in 
partial level densities.  Here we shall show that the coding
becomes quite simple using a real-time formulation of the problem
and the Fast Fourier Transform (FFT).  The 
one-dimensional FFT was first used to calculate level densities by
Berger and Martinot \cite{be74}.  However, for calculating partial
level densities the formulation using the trace of the real-time
Green's function is more
transparent and can be easily implemented by the multidimensional
FFT.  The codes to 
perform the calculations under different conditions are described in
the appendix and are available for download.  

\section{Real-time method}

To derive the equations of the real-time method, 
let us consider a Fock space of $N_{p}$ orbitals with the Hamiltonian
\be
\hat H = \sum_i^{N_p} \varepsilon_i a^\dagger_i a_i
\ee 
where $\varepsilon_i$ are the single-particle energies of the orbitals.
The total level density $\rho(E)$ is defined as
\be
\rho(E) = {\rm Tr} \left( \delta(\hat H - E)\right)
\ee
Here the trace runs over all states of the many-particle Fock space, 
i.e. with any number of particles in the space.
Next, the $\delta$-function is represented by the Fourier
transform
\be
\delta(\hat H - E) ={1\over 2 \pi}\int_{-\infty}^\infty d t  \,e^{i E t}
e^{-i\hat H t}.
\ee
The trace of the operator in this equation is easy to evaluate due to
the independent-particle character of the Hamiltonian. It is
\be
G(t) \equiv {\rm Tr}\left( e^{-i\hat H t}\right) = \prod_i^{N_p} (1 + e^{-i \varepsilon_i t} ).
\ee
In practice, the computation is carried out by building a table of $G$ as a function of $t$ and
then applying the Fourier transform by the FFT algorithm.  Specifically,
we apply the discrete FFT with time points $t_i$ as
\be
\tilde G(E_i) = {\tt FFT}\,(G(2 \pi t_i))
\ee
with $t_i$ forming a mesh with $N_t$ points separated 
by a fixed interval $\Delta t$.  The result is $\tilde G$, an
array with the elements giving the number of levels in an energy
interval $ \Delta E = 1 / \Delta t$ around the point $E_i$,  
\be
\tilde G(E_i) = \int_{E_i}^{E_i+\Delta E} dE\, \rho(E) 
\ee  
To make the
results completely transparent, it is helpful to discretize the
single-particle energy spectrum with the same $\Delta E$.  Provided the 
discretization in $t$ is
sufficiently fine, the output of the FFT will be an integer in each energy bin.

We have coded Eq. (5) and Eq. (8,9) below for the total level density 
partial level densities in {\tt rt\_levels3.py}.  The program is 
described in the Appendix and available
for download.  To illustrate its use, we calculate the neutron level 
density of the heavy deformed nucleus $^{164}$Dy, taking the
single-energies from the Hartree-Fock spectrum calculated with 
the Gogny D1S interaction.  The single-particle space has been truncated
to $N_p = 40$ orbitals, taking the orbitals 
of the $^{164}$Dy ground state closest to the Fermi level.  
The excitation energies in that space range from zero to $~120$ MeV,
and the total number of states is 
$2^{40} = 1.1\times 10^{12}$.  The energy binning is taken as
$\Delta E = 0.2$ MeV.  This implies that
the FFT must be carried out with at least $120/0.2\sim 600$ time points.  
The resulting total level density is shown in Fig. 1 as the open 
circles.
\begin{figure}
\includegraphics[ width=8cm]{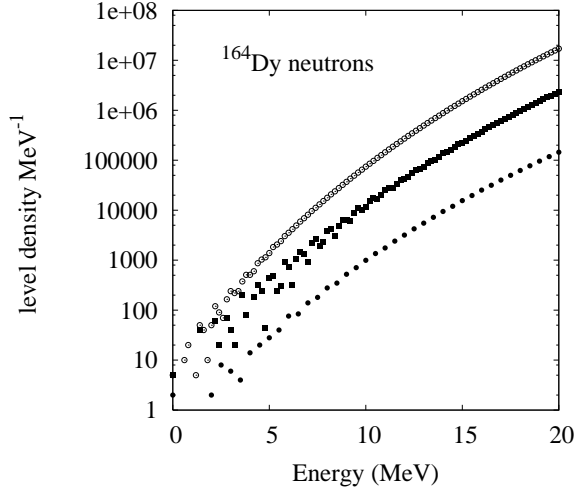}
\caption{Level densities for neutron levels in $^{164}$Dy. Open circles
are the total level densities.  Black squares are projected onto
states with equal numbers of particles and holes.  Black circles are
further projected onto states with $J_z = 0$.  The total and 
number-projected densities are binned with $\Delta E = 0.2$ MeV bins;
the $J_z$-projected densities are with $0.5$ MeV bins.
}
\end{figure}

In practical applications we often would like level densities projected 
onto conserved quantum numbers.  If the quantum numbers are additive,
the projections can also be conveniently 
carried out by Fourier 
transform.  For example, number projection is performed by introducing
a second $\delta$-function in the trace formula,
\be
\delta ( \hat N -N) = {1\over 2 \pi} \int_\infty^\infty d\phi_N\, e^{i \hat N
\phi_N}e^{-iN\phi_N}
\ee
where $\phi_N$ is a gauge angle.  As before, the integral is evaluated
as a discrete Fourier transform.
This method for number projection is in common use, eg. in the
shell-model Monte Carlo treatment of level densities\cite{or94}
and in the extended Hartree-Fock-Bogoliubov theory of ground state
energies\cite{an01}.   The program {\tt rt\_levels3.py}
also includes the coding for the needed two-dimensional FFT, which we
write as
\be
\tilde G(E_i,N) = {\tt FFT2}\,(G(2 \pi t_i, 2\pi \phi_k))
\ee  The discretization in
gauge angle $\phi_k$ require at least as many angles as there are in the range
of $N$ values in the output.  This calculation is illustrated by the
black squares in Fig. 1.
The single-particle 
spectrum is the same as in the previous example; one sees that the
projected level density is as much as an order of magnitude smaller.

The same technique can be used for any other additive quantum number.  
Besides
particle number, we would like to project on $J_z$, 
the $z$-component of angular
momentum.  If the nuclear field is axially symmetric, the orbitals have
a well-defined $J_z$ quantum number and an additional
variable can be added to $G$ corresponding to
rotation angles $\phi_z$ about the $z$-axis. 
The required three-dimensional Fourier transform is 
also coded in {\tt rt\_levels3.py}.  We write it as
\be
\tilde G(E_i,N,J_z) = {\tt FFT3}\,(G(2 \pi t_i, 2\pi \phi_k, 2\pi \phi_z)).
\ee
The result for the combined $N$- and $J_z$-projection of the neutron level 
density in $^{164}$Dy is shown in Fig. 1 as the black circles.

One can apply the same method to the parity operator.  Since
there are only two possible parities, it is sufficient to take
only two angles $\phi=0$ and $\phi=\pi$ in constructing the
$G$ array.  The code {\tt rt\_levels3.py} has the flexibility to 
project on a fixed parity as well as carrying out the $N,Z,J_z$ projections 
at the same time.

One should be aware of two computational issues associated with
the real-time method. First, roundoff error will be come severe
if the size of the many-body space $2^{N_p}$ exceeds the number
of bits in the floating point arithmetic.  The examples in the
Fig. 1 and Fig. 2 below have $N_p=40$, well below the 56
mantissa bits of the double precision arithmetic in the FFT
program library calls. 
Second, the method is only fast if the number of simultaneous
projections is limited. The running time on a laptop is of the
order of seconds or minutes for the one- and the two-dimensional
Fourier transforms.  The three-dimensional transformation is
on the scale of an hour, but higher order transforms would
be quite time-consuming. We describe in the next section an
approximate treatment of projections that might be preferable in
those cases.

\section{Combining combinatorics with statistics}

The most complete decomposition we can envisage here  is to project on proton
number $Z$, neutron number $N$, azimuthal quantum number $J_z$, and parity $P$.
As discussed earlier, parity is easy to include.  But the 4-dimensional
array and Fourier transformed needed to do the $(N,Z,J_z)$ projection is beyond the
scope of laptop computation.  Fortunately, the central limit theorem
allows one to estimate the projections at a factor of 2 in cost for each
projection.  

We illustrate first with a single projection, for example, neutron number
$N$.  The one-dimensional FFT is carried out in the time-energy domain 
with two values of the neutron gauge angle, $\phi_n = 0, \Delta \phi$.  
The angle 
$\Delta \phi$ is chosen to be small enough so that a power series
expansion of $\tilde G$ in that variable is permitted. 
Then we can extract the first and second moments of $\hat N$ for each
bin in $\tilde G(E)$.  Call the total number of states in the bin
$M_E$,
\be
M_E= \tilde G(E,\phi=0).
\ee
The average number of particles and holes in those $M_E$ states is
calculated as
\be
N_1 =\langle \hat N \rangle_E \approx  {{\rm Im }\,  \tilde G(E,\Delta \phi) 
\over M_E \Delta \phi}.
\ee
The mean square number of particles and holes is calculated  as
\be
N_2 = \langle \hat N \hat N \rangle_E \approx {2 \over \Delta \phi^2}\left(
1-{{\rm Re} \tilde G(E,\Delta \phi)
\over M_E}\right).  
\ee
Now treat $N$ as a continuous variable and assume that the distribution in
$N$ is Gaussian, with the same moments $N_1$ and $N_2$.
This gives 
\be
P(N) = { M_E \over \sigma \sqrt{2 \pi}} \exp\left( - {(N-N_1)^2
\over 2\sigma^2}
\right),
\ee
where 
\be
\sigma^2 = N_2-N_1^2.
\ee

The program {\tt rt\_levelsNP.py}  estimates the $N,Z$ and $K$ 
projections using Eqs. (7-10) assuming there are no correlations
between the three variances except for one. Namely, the number parities
of $2K$ and $N+Z$ are always equal, eg.  $2K$ is even if $N+Z$ is even.
So half the of entries in a table of level densities are zero,
and the nonzero ones are on the average twice as large.    
Fig. 2 shows the calculated
levels densities of $^{164}$Dy using {\tt rt\_levelsNP.py}.  
One sees that the three-fold projection has a very strong effect on the 
level density, reducing it by more than two orders of magnitude
in the 5-10 MeV range of excitation energies.
\begin{figure}
\includegraphics[ width=8cm]{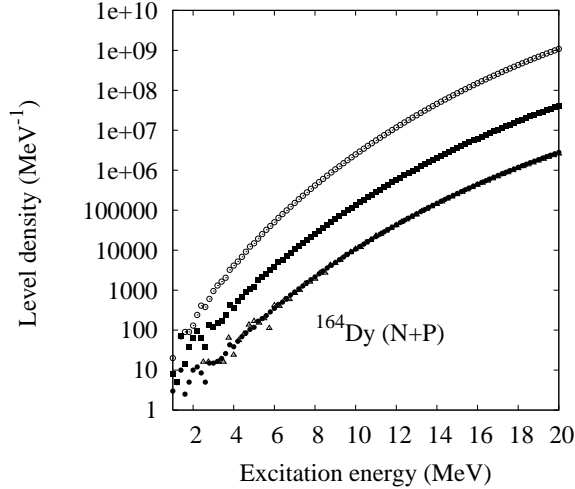}
\caption{\label{np}Level densities for $^{164}$Dy including both neutron and 
proton excitations in a single-particle space totaling 40 orbitals.
Open circles are the total level densities.  Black squares are projected onto
$\Delta N=0$ and $\Delta Z=0$  with respect to the zero-particle
zero-hole ground state, using the approximate projection by 
Eq. 9. 
Black circles are
further projected onto states with $J_z  = 0$ with the same
method.  Also shown with triangles is the exact projection calculated
combining  neutron and proton densities from Eq. (9).  
}
\end{figure}

Also in Fig. 2 we show the exact three-fold projected densities calculated
by folding the neutron and proton level densities obtained from
three-dimensional Fourier transforms.  The results are indistinguishable
for energies over 6 MeV.  The program {\tt foldNP.py} to 
carry out the folding is also included in the package of
codes provided with this article.

\section*{Acknowledgments}  
  
We are grateful for discussions with A. Bulgac, W. 
Nazarewicz, and P.G. Reinhard on the level density problem, 
in the framework of the 2013 INT Program on Large Amplitude Shape Dynamics. 
We would also like to thank Y. Alhassid for a careful reading of the
manuscript.  Support for this work was provided by
the US 
Department of Energy under Grant No. DE-FG02-00ER41132 and 
by MINECO grants Nos. FPA2012-34694, FIS2012-34479
and by the Consolider-Ingenio 2010 program MULTIDARK CSD2009-00064.

\section*{Appendix}

The examples in the text were computed with the programs
{\tt rt\_levels3.py},
{\tt rt\_levelsNP.py}, and {\tt foldNP.py}.
 A tar
file of the three programs together with input data
going with Figs.~1 and 2 is available on  the website
of one of the authors, 
{\tt
www.phys.washington.edu/users/bertsch/computer.html} under
``Level Densities".
They are written in the Python programming language and require
the {\tt numpy} library to run.  The programs have been tested with
version 2.7.3 of Python and 1.6.1 of {\tt numpy}.  

The code for {\tt rt\_levelsNP.py} is reproduced below.
 
\begin{verbatim}
#  rt_levelsNP.py calculates level densities by 1D Fourier transform 
# with neutrons and protons together; N,Z, and K projections are
# calculating assuming that the distributions are Gaussian.
import sys
import math as m
import numpy as np
import numpy.fft as fft

lines = open(sys.argv[1]).readlines()
datafile = lines[0].split()[0]
ss = lines[1].split()
Nt = int(ss[0])
Nphi,Zphi,Kphi = map(float,ss[1:4])
DeltaE = float(lines[2].split()[0])
lines = open(datafile).readlines()
Nnp = len(lines) -2
lines= lines[1:]
print 'Nt, Nnp, DeltaE',Nt, Nnp,DeltaE
print 'phiN,phiZ,phiK', Nphi,Zphi,Kphi
Qvals = []
Evals = np.array([0.0]*Nnp)
etot = 0.0
tau_z = 1
i = 0
for line in lines:
   ss = line.split()
   if len(ss) != 4:
      tau_z = -1
   else:
     K,P,B = map(int,ss[:3])
     if P == 1:  Pex = 0.0;
     if P == -1:  Pex = 1.0
     Qvals.append((K,Pex,B,tau_z))     # K,P,B,tau_z
     E = round(float(ss[3])/DeltaE+1.0e-4)
     Evals[i]= E
     etot += E
     i += 1
print 'DeltaE,etot', DeltaE,etot

# set up the gauged Green's function
ggauged = np.array([[0.0j]*Nt]*4)

Nphase=np.array([0.0]*Nnp)
Zphase=np.array([0.0]*Nnp)
Kphase=np.array([0.0]*Nnp)
nophase = np.array([0.0]*Nnp)
for ip in range(Nnp):
   K,Pex,B,tau_z = Qvals[ip]
   if tau_z == 1:
     Nphase[ip] = Nphi*B
   else:
     Zphase[ip] = Zphi*B
   Kphase[ip] = Kphi*K 
print  

# compute G(t,phi)

G = np.array([0.0j]*Nt)
def make_G(gauge):
  for it  in range(Nt):
    green = 1.0+0.0j
    for ip in range(Nnp):
       E = Evals[ip]  
       exponent  = m.pi*2*E*it/float(Nt)+gauge[ip]
       expgauge = m.e**(1.0j*exponent)
       green = green * (1.0+ expgauge)
    G[it] = green
  return G
 
ggauged[0,:] = make_G(nophase)
ggauged[1,:] = make_G(Nphase)
ggauged[2,:] = make_G(Zphase)
ggauged[3,:] = make_G(Kphase)

#Fourier transform from time to energy

gg_fft = ggauged*0.0
for i in range(4):
   gg_fft[i,:] = fft.fft(ggauged[i,:])/Nt

# extract Gaussian parameters

N0 = np.array([0.0]*Nt); Nsigsq = N0*0.0
Z0 = N0*0.0;  Zsigsq = N0*0.0
K0 = N0*0.0;  Ksigsq = N0*0.0

Nstates = 0.0
for i in range(Nt):
  E = i*DeltaE
  f0 = gg_fft[0,i]
  Ne = f0.real
  Np0 = 1.0; Zp0 = 1.0; Kp0 = 1.0
  if Ne > 0.01:
    N0[i] = gg_fft[1,i].imag/Ne/Nphi
    Z0[i] = gg_fft[2,i].imag/Ne/Zphi
    K0[i] = gg_fft[3,i].imag/Ne/Kphi
    Nsigsq = 2*(1 - gg_fft[1,i].real/Ne)/Nphi**2 - N0[i]**2    
    Zsigsq = 2*(1 - gg_fft[2,i].real/Ne)/Zphi**2 - Z0[i]**2    
    Ksigsq = 2*(1 - gg_fft[3,i].real/Ne)/Kphi**2 - K0[i]**2    
    if Nsigsq > 0.5:
      Nsig = m.sqrt(Nsigsq)
      Np0 = m.e**(-N0[i]**2/(2*Nsigsq))/(2*m.pi)**0.5/Nsig
    if Zsigsq > 0.5:
      Zsig = m.sqrt(Zsigsq)
      Zp0 = m.e**(-Z0[i]**2/(2*Zsigsq))/(2*m.pi)**0.5/Zsig
    if Ksigsq > 0.5:
      Ksig = m.sqrt(Ksigsq)
# Note factor of 2 on line below
      Kp0 = 2*m.e**(-K0[i]**2/(2*Ksigsq))/(2*m.pi)**0.5/Ksig
    Nprojected = Np0*Zp0*Kp0*Ne
    print ' %6.2f %10.1f %6.4f %6.4f %6.4f %10.1f' % (E,Ne,Np0,Zp0,Kp0,Nprojected)
  else:
    print ' %6.2f    0.0' % E   
  Nstates += Ne
print 'Nstates', ggauged[0,0].real,Nstates
\end{verbatim}

\end{document}